\documentclass[twocolumn]{revtex4}
\RequirePackage[dvipsnames,usenames]{color}
\RequirePackage{latexsym,amsmath,amssymb}

\usepackage{graphicx}

\def\lb{\label}
\def\giorno{4/6/2020}
\def\l{\lambda}

\def\a{\alpha}
\def\b{\beta}
\def\ga{\gamma}

\def\s{\sigma}

\def\be#1\ee{\begin{align}#1\end{align}}

\def\^#1{\widehat{#1}}

\def\beql#1{\begin{equation} \label{#1}}
\def\beq{\begin{equation}}
\def\eeq{\end{equation}}

\def\<{\langle}
\def\>{\rangle}

\def\({\left(}
\def\){\right)}

\def\[{\left[}
\def\]{\right]}

\def\eqref#1{(\ref{#1})}

\def\EOR{\hfill $\odot$}

\def\asyref{asy1,asy2,asy3,asy4,asy5,asy6,asy7,asy8,asy9,asy10,asy11}

\begin{document}

\title{Size and timescale  of epidemics in the SIR  framework}

\author{Mariano Cadoni\footnote{mariano.cadoni@ca.infn.it}}
\affiliation{Dipartimento di Fisica, Universit\`a di Cagliari,
Cittadella Universitaria, 09042 Monserrato (Italy); \\ {\rm and} \\
INFN, Sezione di Cagliari, 09042 Monserrato (Italy) }

\author{Giuseppe Gaeta\footnote{giuseppe.gaeta@unimi.it}}
\affiliation{
Dipartimento di Matematica, Universit\`a degli Studi
di Milano, via Saldini 50, 20133 Milano (Italy); \\ {\rm and} \\ SMRI,  00058 Santa Marinella (Italy) }

\date{revised version -- \giorno }

\begin{abstract}
The most important features to assess the severity of an epidemic are its size and its timescale. We discuss these features  in a systematic way in the context of SIR and SIR-type models.  We investigate in detail how the size and timescale  of the epidemic  can  be changed  by acting on the parameters  characterizing   the  model.  Using these results  and  having as guideline the  COVID-19 epidemic in Italy,  we compare  the efficiency  of different   containment strategies  for  contrasting an epidemic diffusion such as social distancing, lockdown, tracing, early detection and  isolation.


\end{abstract}

\maketitle

\section{Introduction}

In contrasting the COVID-19 epidemic, many countries and States resorted to strict social distancing measures, also known as \emph{lockdown}. The general theory of epidemic diffusion grants that this produces a lowering of the epidemic peak and a slow down of the dynamics. While this second aspect can be very positive -- e.g. giving to the Health System the time to strengthen its capacities and more generally to prepare for the epidemic wave -- it also has negative aspects related to the social and economic costs of a lockdown.

The purpose of this paper is twofold: on the one hand, we want to discuss the temporal aspects of a lockdown and more generally of epidemic moderation policies based on social distancing; on the other hand, we want to stress that other tools to counter the epidemic also exist, and they have quite a different impact on the dynamics even in the case they lead to the same reduction of the epidemic peak.

We will mainly work in the framework of the most venerable epidemic model, i.e. the classical (Kermack-McKendrick) \emph{SIR model} \cite{KMK,Murray,Heth, Edel,Britton}; this is specially simple and will allow to obtain some \emph{analytical} results. We will also consider a recently introduced SIR-type model taking into account one of the most striking features of COVID, i.e. the presence of \emph{a large class of asymptomatic infectives} \cite{Crisanti,\asyref} (this is hence called A-SIR model, the ``A'' standing for ``asymptomatic''); in this case we are not able to obtain analytical results, and our discussion is based on numerical computations.

The ideas discussed in this work were formulated in recent works of ours \cite{Cadoni,Gasir,Gavsb}; the discussion, and some of the applications,  given here are original. The applications concern mainly the ongoing COVID epidemic in Italy. A more detailed discussion of this application is presented in a companion paper \cite{CGND}.

The \emph{plan of the paper} is as follows. In Section \ref{sec:sir} we recall the classical SIR model and discuss some of its properties, in particular concerning the temporal aspects of epidemic dynamics. In Section \ref{sec:sir_man} we study how the SIR properties discussed in Sect.~\ref{sec:sir} influence different strategies of epidemic containment of management. This is in a way a preliminary study, as the COVID-19 epidemic has a striking feature, i.e. the presence of a very large set of asymptomatic infectives (or pauci-symptomatic ones). These are difficult to intercept, and thus are active in the epidemic dynamic over a much longer time than symptomatic infectives, which are promptly recognized as such and isolated. We pass then to discuss, in Section \ref{sec:asir}, a SIR-type model recently introduced to take this feature into account \cite{Gasir}, and called A-SIR model, see above. In the following Section \ref{sec:asir_man} we will discuss epidemic management in the framework of the A-SIR model. Our discussion so far is general; in Section \ref{sec:italy} we will apply this general discussion to a concrete case, i.e. the ongoing COVID epidemic in Italy. The A-SIR model describes quite well the epidemiological data so far (provided one takes into account the effect of the restrictive measures put in operation by the Government in two rounds), as shown in \cite{Gasir}; we will discuss some model's projection for the future under different strategies, i.e. changing the model's parameters in different ways -- at first, more sketchily, within the standard SIR framework, and then in more detail within the A-SIR one. Needless to say, these should \emph{not} be seen as forecasts, but as providing a \emph{qualitative} insight into the effects, in particular the temporal ones, of different strategies. Finally, in Section \ref{sec:conclu}, we draw some short Conclusions.

\section{The SIR model}
\label{sec:sir}

The classical SIR model \cite{KMK,Murray,Heth,Edel,Britton} concerns averaged equations for a population of ``equivalent'' individuals (thus in Physics'  language it is a mean field theory); each of them can be in three states, i.e. being Susceptible (of infection), Infected and Infective, and Removed (from the infective dynamic).

The populations of these three classes are denoted as $S(t)$, $I(t)$ and $R(t)$ respectively, and the dynamic is defined by the equations
\begin{eqnarray}
dS/dt &=& - \, \a \, S \, I \nonumber \\
dI/dt &=& \a \, S \, I \ - \ \b \, I \label{eq:sir} \\
dR/dt &=& \b \, I \ . \nonumber \end{eqnarray}
where $\a$ and $\b$ are constant parameters, discussed in a moment.

Note that the third equation amounts to a direct integration, $ R(t)  = R(t_0) + \b \int_{t_0}^t I(y ) d y$.
Moreover, the total population $N$ is constant in time (people dying for the considered illness are considered as removed).

Here the parameter $\a$ corresponds to a \emph{contact rate} (also called infection rate).  More precisely, infection is caused by the meeting of an infective and a susceptible, thus the number of new infected per unit of time is proportional to both $S$ and $I$; the details of the infection process are condensed in the parameter $\a$. This takes into account how easy it is for the pathogen to install in the body of a susceptible once it gets the opportunity to do so (thus what is the immune system  reaction to it), but also the frequency of ``dangerous'' contacts, i.e. of contacts close (or unprotected) enough to give raise to such an opportunity. E.g., for a virus transported by droplets over a certain distance, this will be the frequency  of encounters at less than this distance and without due individual protective devices. \emph{Social distancing measures work on the reduction of $\a$.} Although they cannot act on the biology  of the pathogen  and of the host, they reduce the infection rate  by making it difficult for the pathogen to find  new hosts.

The parameter $\b$ is a \emph{removal rate}; more precisely, it is the inverse of the typical removal time $\tau =\b^{-1}$. For a trivial illness (e.g. a cold) this is essentially the time for the body to heal, but for a dangerous illness removal from the epidemic dynamic will also take place by \emph{isolation}, which is assumed to take place as soon as the infective is recognized as such. Thus, \emph{early detection campaigns}, followed by isolation of infected,  \emph{work on the reduction of $\tau$, i.e. the increase of $\b$}.

The system \eqref{eq:sir} can be solved exactly in the time domain by means of time reparametrization, which allows to linearize the dynamics \cite{Cadoni}.  The solution is given by
\begin{subequations}
\begin{align}
&S=S_0e^{-\a \tau},\label{sol1}\\
&I=I_0+S_0-S_0e^{-\a \tau}-\beta \tau ,\label{sol2}\\
&R=R_0+\beta\tau ,  \label{sol3}
\end{align}
\end{subequations}
where $S_0,I_0$ are initial data and $R_0= N-S_0-I_0$.
The function $\tau(t)$ is defined implicitly by
\beq\lb{ttau}
t=\int_{0}^\tau \frac{d \tau'}{I_0+S_0-S_0e^{-\a \tau'}-\beta \tau'}.
\eeq

\medskip\noindent
{\bf Remark 1.} Exact solutions of  the  SIR model, which are  equivalent  to our  Eqs. (\ref{sol1}),(\ref{sol2}),(\ref{sol3}), (\ref{ttau}) have been derived,  using  a quite different approach by Harko, Lobo and Mak \cite{HLM}.  Instead of performing a time reparametrization,  they first transform the main equation into  a second order equation and then reduce it to a first order   non linear Bernoulli equation, which can be explicitly solved. More recently, Barlow and Weinstein \cite{BaW} obtained an exact solution for the SIR equations in terms of a divergent but asymptotic series \cite{Ba17} (see also \cite{BBGS,Gra} for a different approach to exact solution of SIR and SIR-type models). We thank a Referee for bringing ref. \cite{BaW} to our attention.  \EOR

\medskip\noindent
{\bf Remark 2.} Most of the analysis holding for the standard SIR model can be extended to the case where the infection term takes the generalized form $\a f(S)  I$, with $f(S)$ an arbitrary smooth function -- which in view of its epidemiological meaning should satisfy $f(0)=0$, $f'(0) >0$ \cite{Satsuma}. Some more recent studies, directly related to the COVID epidemic, have also proposed -- explicitly or implicitly -- nonlinear modifications of the bilinear infection term of the standard SIR model \cite{secord,Volpert}. \EOR

\subsection{Epidemic peak and total number of infected}

The most important quantities, which describe the size  of an epidemic are   the \emph{epidemic peak}, i.e. the maximum value $I_*$ attained by $I(t)$ and the \emph{total number infected}   individuals  $R_E$  over the whole time-span of the epidemic.

The value of $I_*$ is crucial  for understanding the maximum pressure that the epidemic is going to put on the health system, whereas $R_E$  gives  a measure of the death toll the epidemic  is going to claim.

It follows immediately from \eqref{eq:sir} that $I(t)$ grows if and only if
\beql{eq:ga} S(t) \ > \ \frac{\b}{\a} \ := \ \ga \ ; \eeq
for this reason $\ga$ is also known as the \emph{epidemic threshold}. An epidemic can start only if $S(t_0) > \ga$, and it stops spontaneously once -- due to the depletion of the $S$ class corresponding to infections -- $S(t)$ falls below $\ga$.

An equivalent way to describe  the epidemic threshold is to introduce the \emph{reproduction number}
$\rho(t)$ :
\beql{eq:rho(t)} \rho(t) \ = \frac{S(t)}{\gamma}, \eeq
which gives  the expected  new  infections generated  by  a single infection; its value at the initial time $t_0$ is also known as the \emph{basic reproduction number}.  The epidemic starts  if $\rho(t_0)>1$, and $I(t)$ attains its peak  value $I_*$ at $t=t_*$,  when $\rho(t_*)=1$.  Containment strategies  aim, by reducing $\a$ and/or by rising $\beta$, to reach  $\rho<1$, thus stopping the epidemic. However,  reaching $\rho<1$ can be  costly in social and economic terms.

\medskip\noindent
{\bf Remark 3.}
A possible alternative containment strategy  is to tune the parameters  $\alpha$ and  $\beta$  in such a way  to maintain $\rho\approx 1$ over a long time-span. In this way the epidemic peak is transformed in huge plateau. In this way a rather unstable  equilibrium  point for the epidemic dynamics is generated, in which   $I$ is held at the constant value $I_*$ for a long time. This may be a useful strategy  if the health system has the capacity to treat $I_*$ infected  until a vaccine  for the disease is developed.  Moreover keeping the infected number at a relatively high level helps in approaching herd immunity and thus preventing a possible second epidemic wave, even if vaccine will not be available. On the other hand, this approach requires to keep the system in an unstable state.  \EOR
\bigskip

By considering the first two equations in \eqref{eq:sir}, we immediately obtain that
\beql{eq:I(t)} I \ = \  I_0 \ + \ (S_0 - S) \ - \ \ga \ \log (S_0 / S) \ . \eeq
This allows to predict immediately the epidemic peak. In fact, we know that this will be obtained when $S=\ga$ (see above), and hence
\beq\lb{ap} I \ = \  I_0 \ + \ (S_0 - \ga) \ - \ \ga \ \log (S_0 / \ga) \ . \eeq
Note that in general in the initial phase of the epidemic $I_0 \simeq 0$; moreover, unless some -- natural or artificial (vaccine) -- immunity is present in the population, $S_0 \simeq N$. This is the case for COVID. Thus we get
\beql{eq:I*} I_* \ = \ N \ - \ \ga \ - \ \ga \ \log (N/\ga) \ . \eeq
Needless to say, these expressions for $I_*$ apply if (and only if) $S_0 > \ga$ and $N>\ga$ respectively: if $S_0 < \ga$ there is no epidemic.

Note that we are able to determine the height of the epidemic peak but, so far, not at which time it is reached, nor more generally any detail about the epidemic development in time. This will be our task in the next subsection.

Before tackling that problem, though, we note that \eqref{eq:I(t)} also characterizes the number of individuals going through the infected state over the whole epidemic period. In fact, at the end of the epidemic we get $I=0$; thus the number $S_E$ of susceptibles at that stage is the (lower) root of the equation (here we use $I_0 \simeq 0$ again)
\beql{eq:Sinf} (S_0 - S_E) \ = \ \ga \ \log (S_0 / S_E) \ . \eeq
This is a transcendental equation and again cannot be solved in closed form, but it is obvious that the solution will depend only on $\ga$, and not on the values of $\a$ and $\b$ producing this value for their ratio. The number of individuals having gone through infection is of course
\beql{eq:Rinf} R_E \ = \ N \ - \ S_E \ . \eeq

In view of the simplicity of the transcendent equation (\ref{eq:Sinf}) for $S_E$, one might look for approximate solutions $S_E(\ga)$. We will not elaborate on this issue.

\subsection{Timescale  of the  epidemic}

There are two main quantities characterizing the timescale of the epidemic:  the \emph{time of occurrence} $t_*$  of the peak and the entire \emph{time-span} $t_E$  of the epidemic, i.e. the time for  which $I(t_E)=0$.  Both $t_*$ and $t_E$   depend on  the parameters  $\a$ and $\b$, and containment measures aimed at reducing the size of the epidemic (i.e. $I_*$ and $R_E$) do in general  have  the effect of increasing  $t_*$ and $t_E$.  We will tackle  this issue in the next subsection by determining how $I_*$, $R_E$, $t_*$ and $t_E$ change by scaling the parameters $\a$ and $\b$.

Let us first  see how $t_*$  can be written in terms  of $\a$ and $\b$.
The time  of occurrence of the peak can be easily calculated using Eqs. \eqref{sol2} and \eqref{ttau}. We have
\begin{eqnarray}
t_*&=&\int_{0}^{\tau_*} \frac{d \tau'}{I_0+S_0-S_0e^{-\a \tau'}-\beta \tau'} \ , \nonumber \\ \tau_*&=&\frac{1}{\a}\log\left(\frac{S_0}{\gamma}\right) \ .
\lb{eq:pt}
\end{eqnarray}
Although here the integral has to be evaluated numerically, the previous expression allows to compute exactly $t_*$ without having to integrate numerically the full system  \eqref{eq:sir}. Moreover, Eqs. \eqref{eq:pt} will allows us to discuss the scaling  of $t_*$  when $\a,\b$  change.

An analytic  expression for $t_*$ can be found using an approximate  solution.     In order to do this we consider the relation between $S$ and $R$; from the first and third equation in \eqref{eq:sir} we easily get
$$ S \ = \ S_0 \ \exp \[ - \, \(\frac{R \, - \, R_0}{\ga} \) \] \ . $$
Using this and recalling \eqref{eq:I(t)}, or equivalently $I(t) = N - S(t) -R(t)$, we can reduce to consider a single ODE, say for $R(t)$. This is written as \beql{eq:R(t)} \frac{dR}{dt} \ = \ \b \ \[ N \ - \ S_0 \,e^{- (R - R_0)/\ga } - R \] \ . \eeq

This is a transcendental equation, and -- albeit the general existence and uniqueness theorem for solutions of ODEs ensures the solution exists and is unique for given initial conditions -- cannot be solved in closed form. It can of course always (and rather easily) be solved numerically.

If -- or until when -- we have $(R-R_0) \ll \ga$, we can expand the exponential in \eqref{eq:R(t)} in a Taylor series, and truncate it at order two. This produces a quadratic equation, more conveniently written in terms of
\beql{eq:P} P(t) \ := \ R(t) \ - \ R_0 \eeq
(hence with initial condition $P_0 = 0$) as
\beql{eq:Rsmall} \frac{dP}{dt} \ = \ \b \ \[ I_0 \ + \ \( \frac{S_0}{\ga} \, - \, 1 \) \, P \ - \ \frac12 \, \frac{S_0}{\ga^2} \, P^2 \] \ . \eeq
This equation can be solved in closed form (see e.g. Sect.~10.2 in \cite{Murray}), yielding
\beql{eq:P(t)} P(t) \ = \ \frac{\a^2}{S_0} \ \[ \( \frac{S_0}{\ga} \, - \, 1 \) \ + \ \kappa \, \tanh \( \frac{\b \,\kappa \, t}{2} \, - \, \phi \) \] \ , \eeq
with constants $\kappa$ and $\phi$ given by
\begin{eqnarray*}
\kappa &=& \sqrt{ \( \frac{S_0}{\ga} \ - \ 1 \)^2 \ + \ \frac{2 \, S_0 \, I_0}{\ga^2} } \ , \\
\phi &=& \frac{1}{\kappa} \ \mathrm{arctanh} \[ \frac{S_0}{\ga} \ - \ 1 \] \ . \end{eqnarray*}
Here again, taking into account that $I_0 \simeq 0$, the expressions are slightly simplified in that we get
\beq \kappa \ = \ \( \frac{S_0}{\ga} \ - \ 1 \) \ . \eeq
Note however that $I_0 \simeq 0$ is a realistic assumption only if we integrate the equations starting from an initial condition at the start of the epidemic; if we have to integrate them from an initial condition when the epidemic is already running, we can not safely assume $I_0 \simeq 0$.

The solution for $R(t)$ is of course obtained by
$$ R(t) \ = \ R_0 \ + \ P(t) \ , $$ see \eqref{eq:P}, which is promptly obtained from \eqref{eq:P(t)}. As $dR/dt = \b I(t)$, the epidemic peak corresponds to the maximum of $R'(t)$, which is the same as the maximum of $P(t)$. The solution \eqref{eq:P(t)} allows to compute the time $t_*$ at which this is attained in a straightforward manner. With standard computations, we get that this is reached at
\beq t \ = \ t_* \ = \ \frac{2 \ \phi}{\b \ \kappa} \ .  \eeq
Recalling now the expressions for $\kappa$ and $\phi$, we get
\beql{eq:t**} t_* \ = \ 2 \ \frac{\mathrm{arctanh} \( \frac{S_0}{\ga} \, - \, 1 \)}{\b \ \[ \( \frac{S_0}{\ga} \, - \, 1 \)^2 \ + \ (2/\ga^2) \, S_0 \, I_0 \]}  \ . \eeq
As usual this gets slightly simpler assuming $I_0 \simeq 0$, which yields
\beql{eq:t*} t_* \ = \ \frac{2}{\b} \ \frac{\mathrm{arctanh} \( \frac{S_0}{\ga} \, - \, 1 \)}{\( \frac{S_0}{\ga} \, - \, 1 \)^2}  \ . \eeq

This (and also the more general \eqref{eq:t**}) shows that $t_*$ is inversely proportional to $\b$. This relation is not surprising, as $\b$ is the inverse of a time (the removal time); in view of this remark, one has to expect that the inverse proportionality holds also without the assumption $(R - R_0) \ll \ga$. This is indeed the case.

The other quantity characterizing the temporal development of the epidemic is  its   entire time-span $t_E$. This   can be computed  by  first setting $I=0$ in Eq.~\eqref{sol2} in order to compute $\tau_E$ and then using  \eqref{ttau} to compute $t_E$ from $\tau_E$. The value of  $\tau_E$ is given by the  (higher)  root of the transcendental equation (we use $I_0=0$  because it is usually very small compared to $S_0$)
\beq\lb{ete}
S_0-S_0e^{-\a \tau_E}-\beta \tau_E=0.
\eeq
$t_E$ is now  obtained by computing  (numerically)  the integral in  Eq.~(\ref{ttau}).

The timescale of the epidemic is obviously related  to its speed. In order to characterize  this speed  we can use the time derivative of the reproduction number $\rho$ evaluated at the peak: $$ V_* \ := \ d\rho/dt|_{t=t{_*}} \ . $$ Using Eqs. \eqref{eq:rho(t)}, \eqref{sol1} and \eqref{eq:pt} we find
\beq\lb{vr}
V_*=  -\a I_*.
\eeq
Notice that $V_*$ is proportional to $\a$; thus, as expected, the epidemic  speed is controlled  by the contact rate $\a$.

\subsection{Scaling properties of the SIR equations}

Our discussion led us to determine, in the approximation $(R-R_0) \ll \ga$, the time at which the epidemic peak is reached. This is surely a useful information, but in many practical cases we would like also to have information (i.e. prediction) about other features of the epidemic curve. In particular, we would like to know when $I(t)$ will descend -- after passing through the peak -- below some safety level $I_s$.

In the $(R-R_0) \ll \ga$ approximation, this information can be obtained from \eqref{eq:R(t)}. But the point is that if we are in the early stages of an epidemic -- as in the present COVID epidemic -- we are not sure that such an approximation will be valid all over the epidemic development. It is thus important to be able to analyze the time behavior of $R(t)$ -- and through it, also of the other quantities $S(t)$ and $I(t)$, see eqs.\eqref{eq:sir}  above -- without the $(R-R_0) \ll \ga$ assumption.

This analysis was performed in a recent paper of ours \cite{Cadoni}, and we will give here the basic step and results emerging out of it.

The first important observation is  that   the amplitude of the peak   $I_*$ \eqref{ap}  and the the total number of infected \eqref{eq:Rinf} $R_E$ are decreasing functions of the parameter $\ga$.   This can be easily shown by  taking the derivatives with respect to $\ga$:
\begin{eqnarray*}
\frac{dI_P}{d\ga}&=& - \log\left(\frac{S_0}{\ga}\right),\\
 \frac{dS_E}{d\ga}&=& - \left(\frac{\ga}{S_E}+1\right)^{-1}\log\left(\frac{S_E}{N}\right).
\end{eqnarray*}
Above the epidemic threshold ($S_0/\ga>1$),  $dI_P/d\ga$ is always negative, whereas  being $S_E<N$, $dS_E/d\ga$ is always positive.
Thus, we have a clear indication of how to fight epidemics and how  to reduce   its  size: we have to increase the epidemic threshold $\ga$, i.e.   we have to scale $\gamma\to \l \ga$ by a factor  $\l>1$.  Increasing  $\ga$ we reduce the reproduction number $\rho$. If we manage  to  reduce $\rho$  below $1$ we simply stop  the epidemic, but even  if we do not manage to go so far we still reduce its severity  by reducing  both  $I_*$ and $R_E$.

However, we have to pay a price. It is quite evident that reducing the reproduction number we slow down the dynamics of the epidemic, so that we increase its time-span, i.e. $t_*$ and $t_E$, and reduce its speed, i.e. $V_*$.

However, one can increase $\gamma$ either by \emph{reducing} $\a$ or by
\emph{increasing} $\b$.  These two different ways of  increasing  $\ga$  may have a different impact of $t_*$ and $t_E$.
That this is true is already evident from the approximate expression  for $t_*$ given  by Eq.~\eqref{eq:t*}: increasing $\beta$ by a factor $\l$, $t_*$ is reduced by a factor $1/\l$, with respect  to the value attained by reducing $\a$.

In order to investigate in a systematic  the scaling properties  of epidemic parameters when we change $\ga\to\l \ga$,  we consider two different  scaling transformations having the same effect on $\ga$: $T^{(1)}$, which reduces  the infection rate and $T^{(2)}$, which increases   the removal rate:
\beq\lb{st}
T^{(1)}:\, \a\to \l^{-1} \a;\quad T^{(2)}: \b\to \l \b,
\eeq
with $\l>1$.

The  quantities $t_*$, $I_*$, $R_E$ and $t_E$   do not transform in a simple way under $T^{(1)}$ and $T^{(2)}$, but  the ratios of  $T^{(1)}$ and $T^{(2)}$-transformed  quantities follow simple scaling laws.

Using the notation $I^{(1)}_P=I_P(\l^{-1}\a)$, $I^{(2)}_P=I_P(\l\b)$ and similarly for the others quantities, to denote rescaled quantities, we have  (see  Ref.\cite{Cadoni} for details)
\begin{eqnarray}
I^{(1)}_* &=& I^{(2)}_* \ , \ \  R^{(1)}_E \ = \ R^{(2)}_E  \ , \nonumber \\
  t^{(2)}_*&=& \l^{-1} \, t^{(1)}_* \ , \ \ t^{(2)}_E \ = \ \l^{-1} \, t^{(1)}_E \ , \lb{resc2} \\
  V^{(2)}_* &=& \l V^{(1)}_* \ . \nonumber
\end{eqnarray}

We see that the  quantities  $I_*$  and $R_E$, describing the size of the epidemic remain invariant (they do not depend on $\a$ and $\b$ separately but only on their ratio  $\ga$), whereas  the quantities $t_*$  and $t_E$ describing the timescale of the epidemic are  reduced by a factor $1/\l$ and the speed  $V_*$ is increased by a factor $\l$.

We stress that our Eqs. \eqref{resc2} entail an important result: different containment measures, which have the same effect on the amplitude of the peak  $I_*$ and  on the total number of infected  $R_E$,  have different impact on the occurrence time  $t_*$ of the peak, on the whole time span of the epidemic $t_E$ and on the epidemic speed $V_*$.

Using  measures that  increase the removal rate $\b$ by a factor $\l$ instead of reducing the infection rate $\a$  by the same factor $\l$ allows to reduce $t_*$ and $t_E$ by a factor $1/\l$. For instance, containment measures  with   $\l=2$  reduce by a half both  the time needed for the epidemic to reach the peak and the whole time-span of the epidemic.  Notice that this epidemic timescale   reduction effect is  more relevant when the reproduction number satisfies $\rho \gg 1$. The  reduction  factor $\l$ is limited by $\l<\rho$ (this is  because for $\l>\rho$ the epidemic does not develop at all).

\medskip\noindent
{\bf Remark 4.}
The scaling of $V_*$ in Eq.~(\ref{resc2}) gives a  simple intuitive explanation of what happens. Acting on $\b$  allows one to speed up the epidemic dynamics while keeping constant the number of infective at the peak and the total number of infected. This is possible because the increasing of the removal rate allows prompt removal of infected individuals. \EOR

\medskip\noindent
{\bf Remark 5.}  Needless to say, the change in timescale can be a positive or a negative outcome -- and thus be sought for, or avoided if possible --  depending on circumstances. If at the beginning of the epidemic the sanitary system is overwhelmed by the number of patients, it is essential to slow down the pace of the epidemic. On the other hand, once the most critical time has gone, it becomes relevant not to be forced to maintain restrictive measures for too long, to avoid huge economical and social costs. This suggests that contrasting the  epidemic should be done with different tools in different phases. \EOR
\bigskip

Summarizing, containment  measures that increase $\b$, e.g. based on tracing and  removal of infected, are more efficient to fight epidemics  in a short time than those that reduce $\a$, e.g. based on social distancing or lockdown. Moreover if by increasing $\b$ we manage to bring $\rho$ below the threshold we simply stop the epidemic, but even if we do not go so far, we can still reduce the size of an epidemic keeping under control its timescale, i.e. without facing the social and economical costs related to a prolonged lockdown or however heavy restrictive measures.

\section{Epidemic management in the SIR framework}
\label{sec:sir_man}

We have so far  supposed that our problem was to analyze the behavior of the SIR system for \emph{given} control parameters $\a ,\b$, and given initial conditions.

When we are faced to a real epidemic, as the ongoing COVID one, any State will try to manage it (which can mean reduce its impact or eradicate it completely, according to what results to be concretely possible -- which depend also on how prompt is the action). When we analyze the situation in terms of the SIR model, this means acting on the parameters $\a$ and/or $\b$.

As discussed above, the strategy are based on two types of actions, i.e. \emph{social distancing}, which can take also the extreme form of a lockdown, and \emph{early detection} (followed by prompt isolation of would-be infected); these impact respectively on the $\a$ and on the $\b$ parameter.

From the discussion of the previous subsections it follows immediately that acting on $\a$ and/or on $\b$ we can reduce the height of the epidemic peak $I_*$ and simultaneously \emph{increase} $t_*$, i.e \emph{slow down the epidemic dynamic}.

This is well known, and indeed one of the reasons for the social distancing measures and lockdown is to slow down the epidemic increase so to have time to prepare for the epidemic wave, e.g. in terms of Hospital -- or Intensive Care Units (ICU) -- capacity, or in terms of Individual Protection Devices (IPD) stocking.

The problem is that social distancing measures -- particularly when they take the form of a lockdown -- are  extremely costly in various ways: e.g. in economical and social terms, as well known; but also in sanitary terms, as a number of pathologies are surely worsened by an extensive time of physical inactivity, and hence a number of extra casualties are to be expected as a consequence of the lockdown. \footnote{It should be stressed that this depends on the lockdown rules; e.g. in Western Europe, albeit lockdown was generally adopted, these were quite different in terms of physical activity from country to country. While in several -- mainly northern -- countries physical exercise and sun exposure were recommended, and other countries' rules were neutral in this respect, Italy and Spain were rather strict in forbidding even individual open-air exercise and all kind of ``not necessary'' (i.e. not work or food shopping related) permanence out of home. This is rather surprising considering that according to the Italian National Institute of Statistics (ISTAT) there are each year about 300,000 deaths due to heart, blood pressure or diabetic illness \cite{istat}, and all of these are surely affected by home confining affected people. It is clear that even a moderate increase, say 10\%, in these rates, could easily result in a death toll surpassing that of COVID, even without considering depressive states surely ensuing \cite{Lancet}, in particular for old people, from long time home isolation and absence of contacts with relatives and friends. Still another problem is the freeze of different kind of cures, e.g. radiotherapy for oncological patients, which has been rather standard in the lockdown.}

Thus, once the first epidemic wave has passed and the Hospital system has been reinforced, it is essential to be able to conclude the lockdown in reasonably short times. Here it should be stressed that albeit in this occasion ``social distancing'' (and hence,in terms of the SIR model, a reduction of $\a$) was considered as equivalent to ``lockdown'', these concepts are \emph{not} the same. E.g., virus transmission rate can be lowered by generalized use of IPD. \footnote{Moreover it is not at all obvious that transmission happens in casual encounters on the street in open air, and not in workplaces or in public transportation, or in shops or at home. Actually all scientific evidence is the other way round, showing that transmission in open air between persons lying at some meters' distance is virtually impossible, while the other environments mentioned above account for most of the transmissions. These are however matters for virologists, and such details are not included in our models; so we will not discuss these matters any more.}


\medskip\noindent
{\bf Remark 6.}
The main result of the previous sections is that containment measures which increase $\b$, such as tracking and removal of infected individuals, are more suitable for fighting epidemics --  in particular once the epidemic peak is on our back, or however if the Hospitals can cope with the number of incoming patients -- in the sense they allow to contain the number of infected people without expanding too much the time span over which  must be maintained.
These kind of strategies are also preferable  -- under the condition mentioned above --  because they have a much smaller social and economic impact than those based on social distancing, in particular if these involve a lockdown. \EOR
\bigskip

\begin{figure}
  \includegraphics[width=200pt]{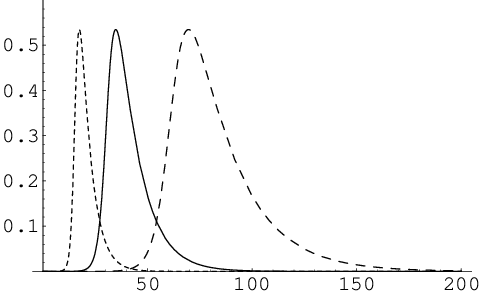}\\
  \caption{Effect of varying parameters in the SIR model while keeping $\ga$ constant. We have considered a population $N=6*10^7$ and integrated the SIR equations with initial conditions $I_0 =10$, $R_0=0$. Setting $\a_0 = 10^{-8}$, $\b_0 =10^{-1} \mathrm{d}^{-1}$, the runs were with $\a = \a_0$, $\b=\b_0$ (solid curve); $\a = \a_0/2$, $\b=\b_0/2$ (dashed curve); and $\a = 2 \a_0$, $\b= 2 \b_0$ (dotted  curve). The curves yield the value of $I(t)/N$, time being measured in days. The epidemic peak reaches the same level, with a rather different dynamics.}\label{fig:sir1}
\end{figure}

\begin{figure}
  \includegraphics[width=200pt]{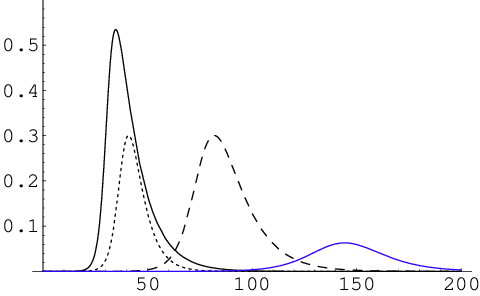}\\
  \caption{Contrasting the epidemic through different strategies. We plot $I(t)/N$ for the same system and initial conditions as in Fig.~\ref{fig:sir1}. Now we consider $\a = \a_0$, $\b=\b_0$ (solid curve, black); $\a = \a_0/2$, $\b=\b_0$ (dashed curve, black); and $\a = \a_0$, $\b= 2 \b_0$ (dotted  curve, black). Actions reducing $\ga$ by the same factor through action on the different parameters produce the same epidemic peak level, but with a substantially different dynamics. We also consider combining the action on the $\a$ and the $\b$ parameters: this is shown in the blue curve, which corresponds to $\a= \a_0/2$, $\b = 2 \b_0$.}\label{fig:sir3}
\end{figure}

In the following we will thus compare the effects of varying the parameters  $\a$ and $\b$ in the SIR model. This will allow us to compare the effects of containment measures based on social distancing with respect to those based on tracing and removal of infected. In particular, we  focus on the temporal aspect, i.e for how long social distancing measures or tracing ones can be needed. In this respect, it is convenient to look at Figure \ref{fig:sir1}, where we show the effect of varying $\a$ ad $\b$ in such a way that $\ga$ is constant; and at Figure \ref{fig:sir3}, where we show the consequences of raising $\ga$ by the same factor (not sufficient to eradicate the epidemic) through action on the different parameters.


It is apparent from (the black curves in) Fig.~\ref{fig:sir3} that the price to pay for a strategy based \emph{uniquely} on social distancing measures -- in whatever way they are implemented -- is that these should be maintained for a very long time. On the other hand, a strategy based on early detection and prompt isolation alone, in agreement with the scaling behavior we have discussed in the previous sections, reduces the peak without slowing down the dynamics. This feature can of course be a positive or negative one, depending on how ready is the Health System to face the epidemic wave.

In real situations, of course, one should act by combining both strategies; different mix of these will produce different time development, and it is not obvious that the choice should be necessarily for the strategy allowing a greater reduction of the epidemic peak, once all kind of sanitary, economic and social considerations are taken into account. This is shown in (the blue curve in) Fig.~\ref{fig:sir3}, where it appears that the strategy combining full action on both parameters reduces greatly the level of the epidemic peak, but also makes the epidemic running for a very long time: this may be convenient or not convenient depending on factors which cannot be taken into account by the purely epidemic model.

For example, we can consider that strict measures should be in effect until the level of infectives descends below a fraction $10^{-4}$ of the population; we denote as $t_s$ the time at which this level is reached. What this means in terms of the duration of the intervention in the different framework considered in Fig.~\ref{fig:sir3} is summarized in Table \ref{tab:sir}.

Note also that here we considered a given initial time of the epidemic and suppose the restrictive measures are maintained until the situation improves enough. In real situations, the restrictive measures will not be set as soon as the epidemic reaches the country, but only after the number of infectives raises above some alert threshold, call this time $t_a$; this has indeed been the pattern in most countries, with some notable exceptions (such as Greece and New Zealand -- they were indeed very lightly struck by COVID). Thus using $t_s$ as an estimate of the length of lockdown leads to overestimating it, and the duration of the measures is better measured by $\tau = t_s - t_a$. This is also considered in Table \ref{tab:sir}, assuming the alert level is the same as the safety one, i.e. $I(t_a) = I(t_s) = 10^{-4} * N$.

In Table \ref{tab:sir} we also give the value of $R_E/N$; this shows clearly that in these simulations we do \emph{not} have $R/\ga \ll 1$ (recall $N>\ga$), hence we can not use the approximation leading to \eqref{eq:t*}. In fact, e.g. in Fig.~\ref{fig:sir3} it appears that raising $\b$ does lead to a (small) delay in the epidemic peak, contrary to what is predicted by the ``small epidemic'' formula \eqref{eq:t*}.

\begin{table}
  \centering
  \bigskip
  \begin{tabular}{|c|c|c||l|r||r|r||r||}
  \hline
  $\a/\a_0$ & $\b/\b_0$ & $\ga/\ga_0$ & $I_*/N$ & $t_*$ & $t_s$ & $\tau$ & $R_E /N$  \\
  \hline
  1 & 1 & 1 & 0.535 & 35 & 125 & 113 & 0.997 \\
  1/2 & 1/2 & 1 & 0.535 & 70 & 251 & 226 & 0.997 \\
  2 & 2 & 1 & 0.535 & 17 & 63 & 56 & 0.997 \\
  \hline
  1 & 2 & 2 & 0.300 & 41 & 94 & 78 & 0.940 \\
  1/2 & 1 & 2 & 0.300 & 82 & 189 & 157 & 0.940 \\
  1/2 & 2 & 4 & 0.063 & 144 & 246 & 182 & 0.583 \\
  \hline
  \end{tabular}
  \caption{SIR model. Height and timing of the epidemic peak $I_*= I(t_*)$ and time for reaching the ``safe'' level ($I(t_s) = 10^{-4} S_0$), together with duration of the interval $\tau = t_s -t_a$ and the fraction $R_E / N$ of individuals having gone through infection, for the different combinations of parameters considered in the numerical runs of Fig.~\ref{fig:sir1} and \ref{fig:sir3}. See text. Note that $R_E/N$ depends only on $\ga$, as guaranteed by eqs. \eqref{eq:Sinf}, \eqref{eq:Rinf}.}\label{tab:sir}
\end{table}

\section{The A-SIR model}
\label{sec:asir}

The SIR model -- besides all the limitations due to its very nature of a ``mean field'' theory, and the absence of any age or geographical structure -- has a weak point when it comes to modeling the COVID epidemic. That is, it does not take into account one of the more striking features of this epidemic, that is the presence of a \emph{large set of asymptomatic infectives}.

\subsection{The model}

In order to take this into account, we have recently proposed a variant of the SIR model, called A-SIR (with the ``A'' standing of course for ``asymptomatic''), see \cite{Gasir}. In this, there are two classes of infected and infective individuals, i.e. symptomatic ones $I$ and asymptomatic ones $J$; to avoid any possible misunderstanding, we stress that we put in the $I$ class all patients which will develop symptoms, also before these arise, and in the $J$ class patients which will not develop symptoms at all. Correspondingly, there are two classes of removed, $R$ for those who are isolated -- and eventually dead or recovered -- after displaying symptoms, and $U$ for those who are not detected as infective by their symptoms and hence are removed only when they naturally heal (unless they are detected by testing contacts of known infectives or by a random test).

The key point is that the mechanism of removal is different for the two classes, and hence so is the removal rate. We denote as $\b$ the removal rate for infectives with symptoms; this corresponds to isolation, with a mean time $\b^{-1}$ from infection to isolation. We denote by $\eta$ the removal rate for asymptomatic infectives, with a mean time $\eta^{-1}$ from infection to healing.

If an individual is infected, it has a probability $\xi \in (0,1)$ to develop symptoms. It is assumed that all infective with symptoms are detected and registered by the Health System; as for asymptomatic ones, only a fraction of them is actually detected (and in this case promptly) isolated.

Within this description, $\xi$ is a biological constant \footnote{We recall our description applies to ``average'' quantities; it is obvious that in a finer description $\xi$ would be depending on the characteristics of the individual who is infected, e.g. his/her age or general health state.}; on the other hand a campaign for detecting asymptomatic infectives -- e.g. by mass random testing -- would result in a reduction of the average time $\eta^{-1}$, i.e. raising $\eta$. On the other hand, a campaign for \emph{tracing contacts} of registered infectives would result in \emph{early isolation} of would-be new infectives -- both symptomatic and asymptomatic -- and hence in a raising of both $\b$ and $\eta$.

It  should be mentioned that a similar model could be considered, where the distinction is not between symptomatic an asymptomatic, but simply between registered and unregistered infective. In this setting $\xi$ can be changed by a testing campaign, while by definition $\eta$ cannot be changed as it corresponds to natural healing of people who are never discovered to be infective; we will not discuss this setting (see \cite{Gavsb} for this point of view).

The A-SIR model is described by the equations
\begin{eqnarray}
dS/dt &=& - \, \a \ S \, (I + J ), \nonumber \\
dI/dt &=& \a \, \xi \, S \, (I +J) \ - \ \b \, I, \nonumber \\
dJ/dt &=& \a \, (1-\xi) \, S \, (I +J) \ - \ \eta \, J, \label{eq:asir} \\
dR/dt &=& \b \, I, \nonumber \\
dU/dt &=& \eta \, J \ . \nonumber \end{eqnarray}
Note that the last two equations amount to integrals, i.e. are solved by $R(t) = R_0 + \b \int_0^t I(y) d y$, $U(t) = U_0 + \eta \int_0^t J(y) \ d y$. Moreover, the total population $N = S+I+J+R+U$ is constant.

Unfortunately, for this model few analytical results are available, and one can only resort to numerical simulations. On the other hand, these show that the A-SIR model can account quite well for the COVID epidemic in Italy \cite{Gasir}, i.e. in the only case where it has been tested against real epidemiological data.

\medskip\noindent
{\bf Remark 7.} Note that according to eqs.\eqref{eq:asir}, symptomatic and asymptomatic patients are infective in \emph{exactly} the same way. This is not completely realistic, and it should be seen as a simplifying assumption. In fact, our goal here is to point out how taking into account the presence of asymptomatic infectives changes the overall picture one would obtain by standard SIR model, and to do this not by a fully realistic model (see e.g. \cite{Fokas,Giordano,NgGui} in this regard) but by the \emph{simplest possible model} taking into account the phenomenon. Actually, it is not clear if the infectivity of asymptomatic should be considered lower or higher than that of symptomatic (recall these are considered infectives only until symptoms become serious enough to grant hospitalization or home isolation, after which they are effectively removed from the infective dynamics): on the one hand the viral charge emanating from them could be lower than for symptomatic, but on the other hand they are as active as fully healthy people, and also people in contact with them do not take the elementary precautions which would be taken if facing somebody with even mild early symptoms (cold, cough, etc). Thus -- in the absence of reliable medical and social information -- we have chose to consider them equally infective to get a simpler model. The problem has been raised in a recent paper by Neves and Guerrero \cite{BrazIta}, where a modified model with a further parameter taking into account the difference in infectivity has been considered. That is, the terms $\a S (I+J)$ of our model are changed into $\a S (I + \mu J)$. These authors state that they have run simulations with different values of $\mu$ and determined that this makes little difference on the overall dynamics. \EOR

\medskip\noindent
{\bf Remark 8.} This seems a suitable place to also point out that infectivity is of course not immediate upon infection, so that an ``Exposed'' class should also be introduced, so to consider a SEIR-type \cite{Murray,Heth,Edel,Britton} model; here too we preferred to disregard this feature, which appears not to be essential when we study the impact of asymptomatic infectives, in order to keep to as simple as possible a model. Similarly, we have considered a constant total population, thus disregarding birth and death processes (in this regard we note that e.g. in the first three months of COVID infection there were about 30,000  cases; in this same period in previous years there were about 300,000 deaths \cite{istat}). \EOR

\medskip\noindent
{\bf Remark 9.} A number of models of SIR type have been proposed in connection with COVID-19. In order to compare model prediction with epidemiological data, some of these models focus on the number of deaths, considered more reliable than those on the infected. As the reader may have  noticed, here instead we do \emph{not} consider at all deaths; we would like to briefly comment on this choice. On the one hand, we do not believe that data about deaths are - at this stage - more reliable than those on infections: data about deaths of people with other pathologies beside COVID are managed in different countries according to different protocols, which within Europe are actually more varied than those leading to classification of infection cases (after PCR or swab tests). Moreover, data about variations in the death rate compared with previous years are not only available with a notable delay, but also mix excess deaths due to COVID with those due to other pathologies (e.g. ictus or heart attacks, which may be increased due to tension and home confinement) and less deaths due to other causes -- such as road or work accidents -- prevented by lockdown. Furthermore,  deaths are classified as due to COVID only in the presence of a test certifying this, and such a test is not always performed.

One should also consider that  mortality is presumably evolving (in a positive sense) as hospitals are less crowded and as doctors learn how to treat COVID cases. In particular, the discovery that many COVID-related deaths are not due to pneumonia but to pulmonary micro-embolism, and the ensuing (rather simple) treatment protocol based on heparin and similar drugs changed in quite a substantial way the medical situation. Moreover, a large part of the deaths in the first wave occurred in senior citizen residences, and were increased by the fact we did not know that asymptomatic infections had such a dominant role in the spreading of infection; now that this is known, and that it has been realized how critical these residences (and in general communities) are, infection of the most fragile part of the population became less frequent, and this also reduces the mortality rate. Thus, in short, we do not believe that one can consider the mortality rate as a constant parameter, and we prefer to deal with the infection data, which are elaborated following a protocol which is possibly not fully reliable, but at least is (for most countries) not changing in time. \EOR

\subsection{The fraction of symptomatic infectives}
\label{sec:xasir}

We will also sometimes write
$$ K(t) \ = \ I(t) \ + J(t) $$ to denote the whole set of infectives; the fraction of symptomatic infectives will thus be
\beql{eq:x} x(t) \ := \ \frac{I(t)}{K(t)} \ , \eeq
while that of asymptomatic will be $y(t) = 1 -x(t)$.

It should be noted that albeit the probability $\xi$ of each new infective to be symptomatic is a constant, the fraction of symptomatic infectives $x(t)$ is a dynamical variable, and will change over time.
The evolution equation governing the change of $x$ is easily written as
\begin{eqnarray*}
\frac{dx}{dt} &=& \frac{d \ }{dt} \ \frac{I}{K} \ = \ \frac{(dI/dt) \, K \ - \ I \, (dK/dt)}{K^2} \\ &=& \frac{\b (\xi S/\ga - 1) I \, K \ - \ I \, (\a S K - \b I - \eta J)}{K^2} \\
&=& \b \,\( \xi \, \frac{S}{\ga} \ - \ 1 \) \frac{I}{K} \ - \ x \, \( \a S \ - \ \b \frac{I}{K} \ - \ \eta \frac{J}{K} \) \\
&=& \b \,\( \xi \, \frac{S}{\ga} \ - \ 1 \) \, x  \ - \ \a S x \ + \ \b \, x^2  \ + \ \eta \, x \,(1 -x) \\
&=& - \, \[ \a \, S (1 -\xi ) \ + \ (\b -\eta) \] \, x \ + \ (\b - \eta) \, x^2  \ . \end{eqnarray*}
Note that here $S$ is itself changing in time, so there is not a nontrivial fixed point for $x(t)$.

\begin{figure}
  \includegraphics[width=200pt]{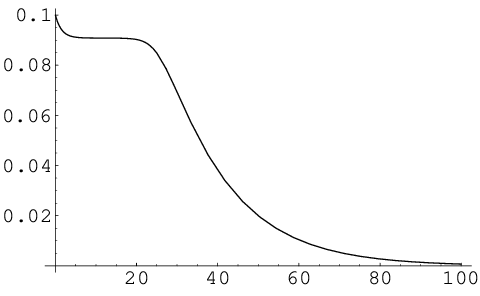}\\
  \includegraphics[width=200pt]{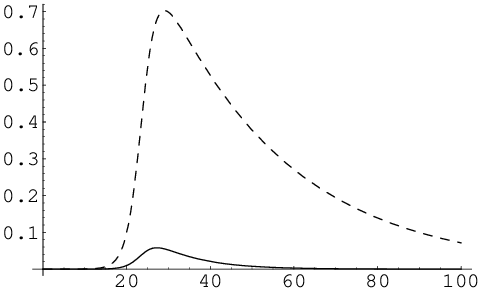}\\
    \caption{Upper plot: the function $x(t)$ computed along the numerical solution to the A-SIR equations \eqref{eq:asir} with $S_0 = 6*10^7$, $\alpha = 10^{-8}$, $\b = 10^{-1}$, $\eta = (1/3)*10^{-1}$, $\xi = 0.1$; the initial conditions are $I_0 =10$, $J_0 = 90$, $R_0 = U_0 = 0$; these imply $x_0 = 0.1$. Note $x(t)$ first goes to a plateau at $x \simeq 0.09$, and then decays  towards zero. Lower plot: the functions $I(t)/S_0$ (solid curve) and $J(t)/S_0$  (dashed curve) for the same numerical solution. It appears that the decay of $x(t)$ starts when $I(t)$ reaches its maximum. This corresponds to $I_*/S_0 \simeq 0.058 $ and is reached at $t_* (I) \simeq 27.24 $; the maximum of $J(t)$ corresponds to $J_*/S_0 \simeq 0.703$ and is reached at $t_* (J) \simeq 29.08 > t_* (I) $.}\label{fig:xxx}
\end{figure}

It should also be noted that in the initial phase of the epidemic $S$ can be considered as constant, $S \simeq S_0 \simeq N$; within this approximation both symptomatic and asymptomatic grow exponentially as a combination of two exponentials, through the linear equations
\begin{eqnarray*}
dI/dt &=& (\a \, \xi \, S_0 \ - \ \b ) \, I \ + \ \a \, \xi \, S_0 \, J \\
dJ/dt &=& \a \, (1-\xi) \, S_0 \, I \ + \ \( \a \, (1 -\xi) \, S_0 \ - \ \eta \) \, J \ .  \end{eqnarray*}
These lead to $x(t) < \xi$; this is easily understood qualitatively: in fact, asymptomatic infectives stay in the infective state for a longer time. This is also shown in Fig.~\ref{fig:xxx}.

\begin{figure}
  \includegraphics[width=200pt]{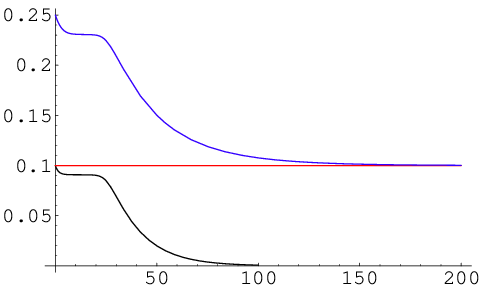}\\
  \caption{For the same numerical solution of the A-SIR equations considered in Fig.~\ref{fig:xxx}, we plot $\chi (t)$ (blue) together with $x(t)$ (black) and the reference line for the value of $\xi$ (red). The function $\chi(t)$ converges to $\xi$ for large $t$, with $x(t) \le \xi \le \chi (t)$ at all times.}\label{fig:xru}
\end{figure}

The situation is just opposite if we look at $R(t)$ and $U(t)$: the fraction of registered removed $\chi = R/(R+U)$ will be higher than $\xi$ at all times, but will converge to $\xi$ for large $t$; see Fig.~\ref{fig:xru}.

Note that this figure offers an explanation for the widely different estimates of $\xi$ appearing in the literature as result of testing campaign. In fact, in several cases it has been attempted to measure $\xi$ from the statistics of testing campaign over a complete population (e.g. for the early evacuation flights from Wuhan \cite{evac}, for the Diamond Princess cruise ship \cite{ship}, or for the V\`o Euganeo community \cite{Crisanti}) or on a random sample of population; but Fig.~\ref{fig:xru} shows that we would obtain quite different results if attempting to measure the fraction of asymptomatic active infectives (which would be done by nasopharyngeal swab testing) or of those who have had the infection (which would be done by serological testing). This remark is specially relevant in view of the planned wide testing campaigns.

\subsection {Epidemic peaks in  the A-SIR model}

One striking feature of the A-SIR model is that  the epidemic threshold is different for the $I$ and the $J$ classes. In fact, it follows from \eqref{eq:asir} that $I$ grows for
\beq \lb{g1}S \ > \ \frac{\b}{\a} \ \frac{x}{\xi} \ = \ \frac{x}{\xi} \ \ga \ := \ \ga_1 \ , \eeq
while $J$ grows for
\beq\lb{g2} S \ > \ \frac{\eta}{\a} \ \frac{1-x}{1-\xi} \ := \ \ga_2 \ . \eeq
This implies that in general the number of symptomatic  and asymptomatic infectives peak at different times.

Let us now show that when $\b \gg \eta$ the peak of  $I$, i.e. $\ga_1$, occurs  before the peak of $J$, i.e. $\ga_2$. In order to do this we use the first equation in \eqref{eq:asir} to eliminate $dt$ from the second and third equations:
\begin{eqnarray*}\lb{eq:asir1}
\frac{dI}{dS}&=&-\xi + \frac{\b}{\a} \frac{x}{S},\\
\frac{dJ}{dS}&=&-(1-\xi) + \frac{\eta}{\a} \frac{1-x}{S}.\\
\end{eqnarray*}

We can now eliminate $x$ from the previous equations by constructing
an appropriate linear combination of them.
We obtain in this way
\beq\lb{fi}
\eta\frac{dI}{dS}+\b \frac{dJ}{dS}=-\omega + \frac{\eta\beta}{\a S},
\eeq
where $\omega:= \eta\xi+\b(1-\xi)$.

\medskip\noindent
{\bf Remark 10.} We note in passing that equation \eqref{fi}  can be easily integrated to give $J$ as  a function of $I$ and $S$:
\beq\lb{eq:J}
J(S)= J_0-\frac{\eta}{\b}(I(S)-I_0) - \frac{\omega}{\b}(S-S_0)+ \frac{\eta}{\a}\ln\left(\frac{S}{S_0}\right).
\eeq
In turn, this equation can be used to eliminate  $J$ from the system   \eqref{eq:asir} and to obtain  an  autonomous, first order, ODE for  $I(S)$. Unfortunately, this equation is an ugly non linear ODE which  cannot be  solved analytically. \EOR
\bigskip

Coming back to the determination of the peaks, we consider the peak for $J$, occurring at $S=\ga_2$, and insert the condition $dJ/dS=0$ into equation  \eqref{fi}. Because  we have $\ga_1 \ge \ga_2$  whenever $S=\ga_2$ belongs to the region of decreasing  $I$,
i.e. to the region where $dI/dS\ge 0$, from the resulting equation we find that   $\ga_1\ge\ga_2$ for $\ga_2\le (\b\eta)/(\omega\a)$. This gives, using the definition of $\ga_2$ in Eq.~\eqref{g1},
\beq\lb{eq:cond}
  \frac{I}{J}\ge \frac{\eta \xi}{\b(1-\xi}.
  \eeq
For values of $S$ in the region   $S \gg \ga_1,\ga_2$, i.e. at small times,  the second  ($S$-dependent) terms in  Eqs. \eqref{eq:asir1}  can be neglected and the early dynamics is determined solely by $\xi$, so that the latter equation gives
 \beq\lb{eq:IJ}
 \frac{I}{J}\approx \frac{\xi}{1-\xi}.
 \eeq

Inserting \eqref{eq:IJ} into \eqref{eq:cond}, we find that in this region the condition \eqref{eq:cond} is satisfied whenever $\b>\eta$.  We have seen in the previous  subsection that $x(t)$, hence also $I/J$,  stays almost constant  before it reaches the peaks for $I$ and $J$ (see the plateau in Figs.~\ref{fig:xxx} and \ref{fig:xru} ). This implies that at least for $\beta\gg \eta$,  the inequality \eqref{eq:cond} remains true until the peaks are reached, hence that $\ga_1>\ga_2$.

\section{Epidemic management in the A-SIR framework}
\label{sec:asir_man}

We have seen above that in the frame of the standard SIR model the epidemic peak can be reduced in several ways, i.e. acting on one or the other (or both) of the parameters $\a$  and  $\b$; these different strategies, however, yield quite different results in terms of the temporal development of the epidemic.

It is quite natural to expect that the same happens in the frame of the A-SIR model. But, as mentioned above, we are not able so far to provide analytical results for the A-SIR model; we will thus have to resort to numerical integration.

It should be mentioned that here we have one further parameter on which we have no control, i.e. $\xi$, and one further parameter -- besides $\a$ and $\b$ -- on which we do have control, i.e. $\eta$.

\medskip\noindent
{\bf Remark 11.} We have seen in Sect.~\ref{sec:asir} that the epidemic threshold is different for the $I$ and the $J$ classes.
Moreover we have shown that, at least for $\beta \gg \eta$, the peak $\ga_1$ for $I$ occurs before  (in time) that for $J$, i.e. $\ga_2$;  recalling that $dS/dt < 0$, this means $\ga_1 > \ga_2$. This fact has an intuitive explanation. At very early times the dynamics is ruled by $\xi$ and we have  $x\approx\xi$, which implies $I/J\approx \xi/(1-\xi)$.  Because $\ga_1$ is proportional to $\beta$ and $\ga_2$ is proportional to $\eta$, for $\beta \gg \eta$ we have in this region $\ga_1>\ga_2$ (see Eqs. \eqref{g1} and \eqref{g2}).  At intermediate times, although  $x$ changes and we have $x<\xi$, the difference in the removal rates $\beta,\eta$ does not affect fully the dynamics, and $x$ stays almost frozen in a plateau till the peaks are reached, preserving in this way the $\ga_1>\ga_2$  relation.
It is only at late times, that is after surpassing the peaks, that the difference in the removal rates $\beta,\eta$ affects fully the dynamics and correspondingly $x$ changes significantly. \EOR
\bigskip

We can thus act on our system in three independent ways, i.e. by \emph{social distancing} (reducing $\a$), by \emph{early isolation} of symptomatic (raising $\b$), and by \emph{detection of asymptomatic} (raising $\eta$).

It should be stressed that while implementation of the first strategy is rather clear, at least in principles (things are less clear when other, e.g. economical and social, considerations come into play), for the other two strategies the implementation is less clear, and in practice consists of two intertwined actions: \emph{tracing contacts} of known infectives, and \emph{large scale testing}. Tracing contacts allows to identify those who are most probably going to be the next infectives; they can then be isolated \emph{before} any symptom shows up, i.e. $\b^{-1}$ can be reduced below the incubation time. On the other hand, this strategy requires to test would-be infectives after some reasonable delay, to avoid quarantining a huge set of non-infective individuals. One could also imagine, if tests were widely available, that specific classes of citizens (e.g.sanitary operators, or all those being in contact with a large number of persons) could be tested thus identifying a number of asymptomatic infectives not known to have had contacts with known infected; this would lead to isolation of asymptomatic, and hence to a raising in $\eta$. Note $\eta$ is also raised by tracing contacts, as many of the infected found in this way would be asymptomatic, but would be promptly isolated in this way.

To put things in a simple way, \emph{actions on $\b$ and on $\eta$ go usually, in practice, together}. Thus we should essentially distinguish between social distancing on the one side, and other tools on the other side.

\begin{figure}
  \includegraphics[width=200pt]{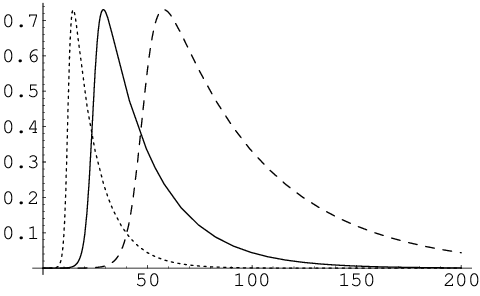}\\
  \includegraphics[width=200pt]{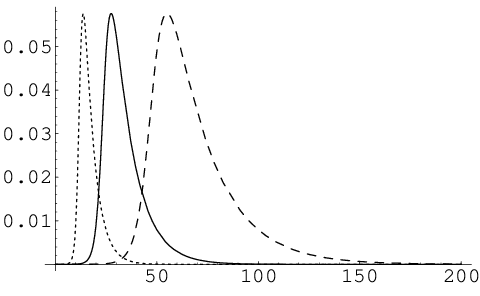}\\
  \caption{Effect of varying parameters in the A-SIR model while keeping $\ga$ constant. We have considered a population $N=6*10^7$ and integrated the A-SIR equations with $\xi = 0.1$ and initial conditions $I_0 =10$, $J_0 = 90$, $R_0=U_0=0$. Setting $\a_0 = 10^{-8}$, $\b_0 =10^{-1} \mathrm{d}^{-1}$, $\eta_0 =4*10^{-2} \mathrm{d}^{-1}$, the runs were with $\a = \a_0$, $\b=\b_0$, $\eta=\eta_0$ (solid curve); $\a = \a_0/2$, $\b=\b_0/2$, $\eta=\eta_0/2$ (dashed curve); and $\a = 2 \a_0$, $\b= 2 \b_0$, $\eta= 2 \eta_0$ (dotted  curve). In the upper plot, the curves yield the value of $K(t)/N$, time being measured in days; in the lower plot, thy refer to $I(t)/N$ . The epidemic peak reaches the same level, with a rather different dynamics.}\label{fig:asir1}
\end{figure}

\begin{figure}
  \includegraphics[width=200pt]{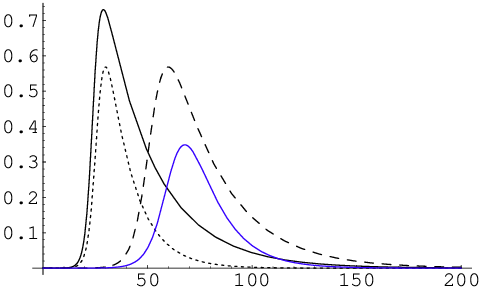}\\
  \includegraphics[width=200pt]{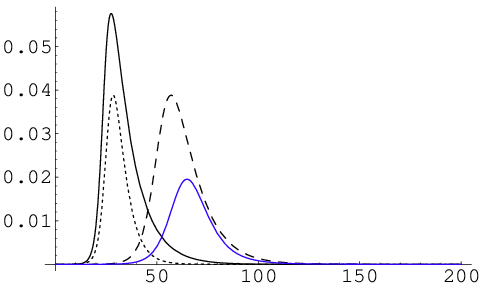}\\
  \caption{Contrasting the epidemic through different strategies. We plot $K(t)/N$ (upper plot) and  $I(t)/N$ (lower plot) for the same system and initial conditions as in Fig.~\ref{fig:sir1}. Now we consider $\a = \a_0$, $\b=\b_0$, $\eta=\eta_0$ (solid curve, black); $\a = \a_0/2$, $\b=\b_0$, $\eta=\eta_0$ (dashed curve); $\a = \a_0$, $\b= 2 \b_0$, $\eta=2 \eta_0$  (dotted  curve); and $\a = \a_0/2$, $\b=2 \b_0$, $\eta=2 \eta_0$ (solid curve, blue). Actions reducing $\ga$ by the same factor through action on the different parameters produce the same epidemic peak level, but with a substantially different dynamics.}\label{fig:asir2}
\end{figure}

In order to do this, we resort to numerical computations for a system with parameters strongly related to those of the SIR numerical computations considered in Sect.~\ref{sec:sir_man} above, and along the same lines. The results of these are illustrated in Figs.~\ref{fig:asir1} and \ref{fig:asir2}; these should be compared with Figs.~\ref{fig:sir1} and \ref{fig:sir3}.

We also consider the quantities analogous to those considered in Table \ref{tab:sir}; these are given in Table \ref{tab:asir}.

\begin{table}
  \centering
  \bigskip
  \begin{tabular}{|c|c|c|c||l|l|r||r|r|r||}
  \hline
  $\a/\a_0$ & $\b/\b_0$ & $\eta/\eta_0$ & $\ga/\ga_0$ & $K_*/N$ & $I_*/N$ & $t_1$ & $t_2$ & $t_s$ & $\tau$  \\
  \hline
  1   & 1   & 1   & 1 & 0.730 & 0.058 & 29 & 28 &  94 &  82  \\
  1/2 & 1/2 & 1/2 & 1 & 0.730 & 0.058 & 58 & 55 & 188 & 164  \\
  2   & 2   & 2   & 1 & 0.730 & 0.058 & 14 & 14 &  47 &  41  \\
  \hline
  1/2 & 1   & 1   & 2 & 0.568 & 0.039 & 60 & 57 & 129 & 104  \\
  1   & 2   & 2   & 2 & 0.568 & 0.039 & 30 & 29 &  65 &  52  \\
  1/2 & 2   & 2   & 4 & 0.348 & 0.020 & 68 & 65 & 124 &  92  \\
  \hline
  \end{tabular}
  \caption{A-SIR model. Height and timing of the epidemic peak $K_* [t_1]$ and of the peak for symptomatic $I_*=I[t_2]$ (note $t_1 \not= t_2$) and time for reaching the ``safe'' level ($I(t_s) = 10^{-4} S_0$), together with duration of the interval $\tau = t_s -t_a$ defined in terms of symptomatic infectives, for the different combinations of parameters considered in the numerical runs of Fig.~\ref{fig:asir1} and \ref{fig:asir2}. See text. In all cases nearly all the population goes through infection, and a fraction $\approx \xi$ with symptoms.}\label{tab:asir}
\end{table}

It is rather clear that the same general phenomenon observed in the SIR framework is also displayed by the A-SIR equations. That is, acting only through social distancing leads to a lowering of the epidemic peak but also to a general slowing down of the dynamics, which means restrictive measures and lockdown should be kept for a very long time. On the other hand, less rough actions such as tracing contacts and the ensuing early isolation allow to reduce the peak without having to increase the time length of the critical phase and hence of restrictions.

\section{The ongoing COVID epidemic -- Italy}
\label{sec:italy}

Our considerations were so far quite general, and in the numerical computations considered so far we used parameter values which are realistic but do not refer to any concrete situation.

It is quite natural to wonder how these considerations would apply in a concrete case; we will thus consider the case we are more familiar with, i.e. the ongoing COVID epidemic in Italy. This is also of more general interest since Italy was the first Western country to be heavily struck by COVID; as it appears Asian countries have dealt with the problem in a rather different way than it is done in the West, the Italian case is used as a test case in many other countries to foresee what can be the developments taking one or another choice. \footnote{This is even more true considering that the Italian Health system is on regional basis, and different regions made different choices -- with quite different results in terms of mortality and virus diffusion. We will not enter into such details.}

Our data and considerations refer to the moment of closing the first submitted version of this paper, i.e. April 25, 2020. Time is measured in days, with Day 0 being February 20, 2020.

The problem in analyzing a real situation is that the parameters are not constant: people get scared and adopt more conservative attitudes, and government impose restrictive measures. All these impact mainly on \emph{social distancing}, i.e. on the contact rate $\a$ \footnote{There are exceptions to this rule: e.g. in Veneto the regional strategy has been more focusing on early detection and prompt isolation, also thanks to contact tracing; with quite good results \cite{Crisanti}.}. Moreover, any measure shows its effect only after some delay, and of course not in a sharp way -- as incubation time is not a sharp constant but rather varies from individual to individual.

\subsection{Fit of real data by the SIR and A-SIR dynamics}

In related works \cite{Cadoni,Gasir} we have considered the SIR and the A-SIR models, and estimated parameters values -- that is, $\a$ and $\b$ for the SIR model, $\a, \b ,\eta$ and $\xi$ for the A-SIR one -- allowing them to satisfactorily describe the first phase of the COVID epidemic in Italy; we have then assumed that each set of restrictive measures has the effect of changing $\a$ to $r \a$ (with $0 < r < 1$ a reduction factor), and that this effect shows on after a time $\b^{-1}$ from the introduction of measures. This is a  very rough way to proceed, but it is in line with the simple approach of SIR-type modeling.

In Italy (total population $N \simeq 6*10^7$) a first set of restrictive measures all over the national territory was adopted on March 8, and another one on March 22. In the framework of the A-SIR model, our estimate of the parameters for the first phase of the epidemic (that is, before the first set of measures could have any effect, i.e. before March 15) was the following, with time measured in days:
\begin{eqnarray} \a_0 & \simeq & 3.77 * 10^{-9} \ , \ \ \b^{-1} \ \simeq \ 7 \ , \ \ \eta^{-1} \ \simeq \ 21 \ ; \nonumber \\ \xi & \simeq & 1/10 \ . \end{eqnarray}
See \cite{Gasir} for details on how these estimates are obtained.

When looking at data in the following time, we assumed that the contact rate $\a$ is reduced by the restrictive measures (and public awareness) and is changed from $\a_0$  into $\a_1 =r_1 \a_0$ between March 15 and March 29, and then into $\a_2 = r_2 \a_0$ from March 29 to April 25. Our best fit for the reduction factors $r_i$ is
$$ r_1 \ = \ 0.5 \ ; \ \ r_2 \ = \ 0.2 \ . $$

\begin{figure}
  \includegraphics[width=200pt]{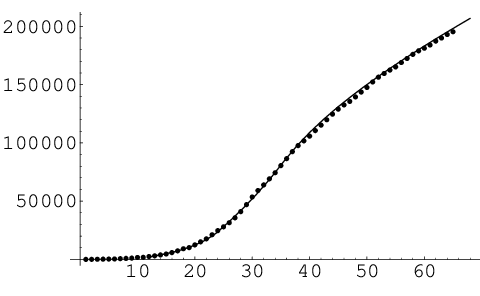}\\
  \caption{Fit of epidemiological data (points) by the function $R(t)$ arising from numerical solution of the A-SIR equations; see text for parameter values and other details. Here time is measured in days, with day 0 being February 20. }\label{fig:ITA}
\end{figure}

\begin{figure}
  \begin{tabular}{cc}
  \includegraphics[width=120pt]{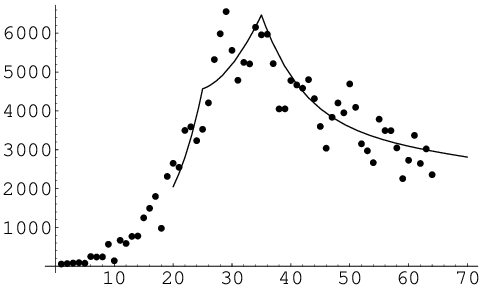} &
  \includegraphics[width=120pt]{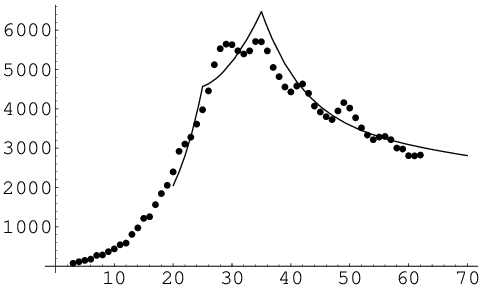} \end{tabular}
  \caption{Fit of epidemiological data (points) for new daily infectives, obtained by the function $I(t)$ arising from numerical solution of the A-SIR equations; see text for parameter values and other details. Here time is measured in days, with day 0 being February 20. Left: Raw data for daily new infectives. Right: Smoothed data, smoothing corresponding to average over five days. }\label{fig:ITAINC}

\end{figure}

In Fig.~\ref{fig:ITA} we plot epidemiological data for the cumulative number of detected infections against the numerical solutions to the A-SIR equations for these values of the parameters, and for initial data obtained also from the analysis of real data. This shows a rather good agreement. See also Fig.~\ref{fig:ITAINC}.

If we were considering the SIR model, some care should be put on the estimation of $\b$; to avoid any confusion, we will here denote by $B$ the removal rate for the SIR model, keeping $\b$ for the removal rate of symptomatic infectives in the A-SIR model.

In the SIR model, indeed, there is no distinction between symptomatic and asymptomatic infectives; thus $B$ is an average quantity computed over the whole set of infectives. This means that it  is related to the A-SIR quantities by
\beq B \ = \ \b \, x \ + \ \eta \, (1-x) \ . \eeq

As we have seen above in Sect.~\ref{sec:xasir}, $x=x(t)$ is a dynamical variable. Actually $x(t) < \xi$, but it remains sufficiently near to $\xi$ in the growing phase of the epidemic. So our rough estimate for $B$ will be
\beq B \ \simeq \ \b \, \xi \ + \ \eta \, (1-\xi ) \ ; \eeq
with the values given above for the A-SIR parameters, this yields
\beql{eq:BBB} B \ \simeq \ 2/35 \ \simeq \ 0.057 \ ; \ \ B^{-1} \ \simeq \ 17.5 \ . \eeq This number by itself, compared with the incubation time of about 5 days for COVID, urges for a campaign of early tracing.

\begin{figure}
  \includegraphics[width=200pt]{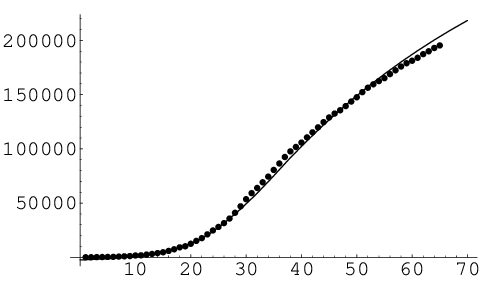}\\
  \caption{Fit of the epidemiological data for Italy by the SIR model with varying $\a$ parameter. This plot reports epidemiological data for the cumulative number of registered infective, thus $R(t)$, day by day; day zero is February 20. The fitting curve is the solution to the SIR equations with initial values (computed in day 14, i.e. March 5) $I_0 = 16,914$, $R_0 = 3,862$ and for a population of $S_0= 6*10^7$. The parameters value were $\a_0 = 3.1*10^{-9}$ before March 15, then changed into $\a_1 =r_1 \a_0$ between March 15 and March 29, and then into $\a_2 = r_2 \a_0$ from March 29 on; and $\b = 2/35 \simeq 0.057$ as stated by Eq.~\eqref{eq:BBB}.  }\label{fig:ITAsir}
\end{figure}

We can then try to compare a numerical solution of the SIR equations with these parameters -- and similar reduction factor for the contact rate as a result of restrictive measures -- with epidemiological data.  This is shown in Fig.~\ref{fig:ITAsir}, showing a reasonable agreement. Note however that in order to obtain this agreement we had to set
\beq \a_0 \ = \ 3.1 * 10^{-9} \ . \eeq

We note that Figures \ref{fig:ITA}, \ref{fig:ITAINC} and \ref{fig:ITAsir} are reproduced from {\tt arxiv:2003.08720v3}; the published version of this paper \cite{Gasir} contains updated plots, showing that the model describes correctly the epidemiological data up to date of publication, provided one takes into account another step in the reduction factor $r$; see \cite{Gasir} for details. 

\subsection{Simulating further interventions. The SIR picture}

Albeit we believe that the A-SIR dynamics is better suited to describe the COVID epidemics, we will start by considering simulations of further interventions on the epidemic dynamic within the SIR framework; this with the purpose so that the reader can observe how the same qualitative picture we will find -- and study in more detail -- within the A-SIR framework is also rendered by this more classical model.

\begin{figure}
  \includegraphics[width=200pt]{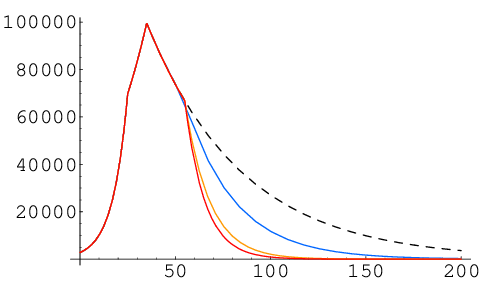}\\
  \caption{Simulation of different containment strategies (after day 55, for Italy) within the SIR model; the initial parameters are $\a_0 = 3.1*10^{-9}$, $\b_0=2/35$, with $\a$ changing to $\a_1 = \a_0/2$ and to $\a_2 = \a_0/5$ at days 25 and 35 respectively, see text. After day 55 we consider a dynamics going on with the same parameters (black curve, dashed) together with: one in which the parameter $\a$ is further halved, thus getting to $\a_3=\a_0/10$, while $\b$ remains at $\b = \b_0$ (blue); one with $\a$ staying at $\a= \a_2$ while $\b$ is doubled, thus going at $\b = 2*\b_0$ (orange); and one where $\a$ is halved and $\b$ is doubled (red). The plot shows the function $I(t)$ along the numerical solutions.}\label{fig:SIR_man}
\end{figure}

In Fig.\ref{fig:SIR_man} we have considered an epidemic dynamic which until day 55 is just the one considered in Fig.\ref{fig:ITAsir}, while in subsequent days is either continuing in the same way, either has a further halving of the $\a$ parameter, either has a doubling of the $\b$ parameter, either has both a halving of $\a$ and a doubling of $\b$. It is rather clear from the picture that acting on $\b$ is much more effective from the point of view of shortening the time needed to get to a low level of infectives in the population. Moreover, it also appears that once $\b$ is increased, an accompanying cut in $\a$ (i.e. further social distancing measures) is nearly irrelevant from this point of view -- while we know it would be very hard in terms of economic and social costs.

The more detailed analysis in terms of the A-SIR model, to be conducted in the next subsection, will confirm this picture.

\subsection{Simulating further interventions. The A-SIR picture}

We can now simulate further intervention, with four possible strategies:
\begin{enumerate}

\item Do nothing, i.e. keep $\a,\b,\eta$ at the present values;

\item Further reduction of $\a$  by a factor $\s$;

\item Increase of $\b$ and $\eta$ by a factor $\s$;

\item Simultaneous reduction of $\a$ and increase of $\b$ and $\eta$, all by a factor $\s$.

\end{enumerate}

We stress that these are not equally easy; we have considered the same reduction/increase factor in order to better compare the outcome of these three actions, but as so far all the intervention has been on $\a$, one should expect that further reducing it is very hard (even more so considering ``side effects'' of this on society, economics, and other sanitary issues); on the other hand, no specific campaign designed to increase $\b$ has been conducted nationwide so far \footnote{With the exception of Veneto, as already recalled in a previous footnote.}, so we expect action in this direction would be quite simpler an has more room for attaining a significant factor.

Despite these practical considerations, as already announced, we consider the same factor for the reduction of the contact rate and for the increase of the removal rates, in order to have a more direct comparison of the effects of these strategies.

\begin{figure}
  \begin{tabular}{|c|}
  \hline
  \includegraphics[width=200pt]{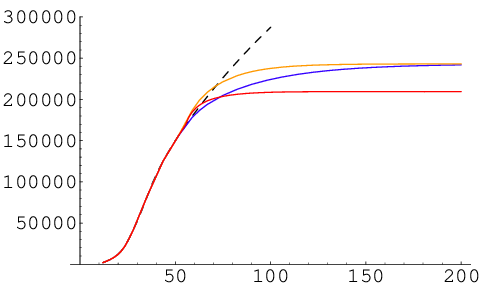}\\
  $R(t)$ \\
  \hline
  \includegraphics[width=200pt]{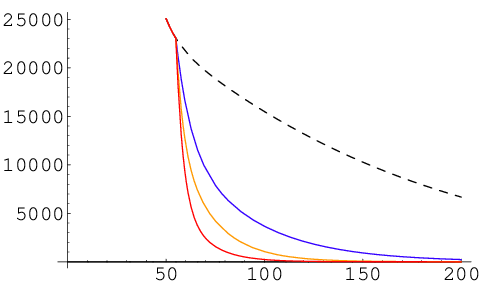}\\
  $I(t)$ \\
    \hline
  \includegraphics[width=200pt]{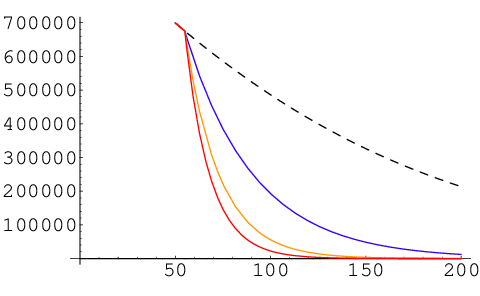}\\
  $K(t)$\\
    \hline
  \includegraphics[width=200pt]{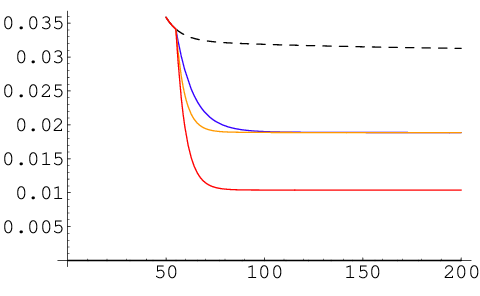}\\
  $x(t)$\\
    \hline \end{tabular}
  \caption{Simulation of different strategies after day 55 for Italy. Strategies correspond to those listed in the text, with factor $\s = 2$. The curves illustrate the predicted outcome under different strategies: no action (black, dashed); further reduction of $\a$ (blue); increase of $\b$ and $\eta$ (yellow); reduction of $\a$ and increase of $\b$ and $\eta$ (red). The plots represents, from the upper one to the lower one: cumulative number of symptomatic infected $R(t)$; number of symptomatic infectives $I(t)$; total number of infectives $K(t)= I(t)+J(t)$; ratio of symptomatic to total infectives, $x(t) = I(t)/K(t)$.}\label{fig:strat}
\end{figure}

We have then ran numerical simulations corresponding to the different strategies listed above; the outcome of these is given in Fig.~\ref{fig:strat}. Note that within each strategy the decay of $I(t)$ is faster than that of $K(t)$, and that the ratio between $I(t)$ ad $K(t)$, i.e. $x(t)$, is in all cases rather far from $\xi = 0.1$, as discussed in Sect.~\ref{sec:xasir} above.

We also report in Table \ref{tab:meas} the relevant expected data for the time $t_s$ at which a safety level $I_s$ -- now assumed to be $I_s = 3,000$, i.e. half of the level reached when the first restrictive measures were taken -- is reached, the time interval $\tau = t_s -55$ from the day the new strategy is applied, and the total count $R_E$ of symptomatic infected. This table also reports the asymptotic value $x_E$ of the ratio $x(t)=I(t)/K(t)$ of symptomatic to total infectives; this is relevant in two ways: one the one hand one expects that only symptomatic infected may need hospital care, so a low level of $x$ means that albeit a number of infections will still be around only a small fraction of these will need medical attention, and COVID will not absorb a relevant part of the Health System resources; on the other hand, this also means that most of the infectives will be asymptomatic, i.e. attention to identifying and isolating them should be kept also when the number of symptomatic infections is very low. Note that for the strategy (1) what we report is not really an asymptotic value, but the value expected at December 31, as the decay is too slow.

\medskip\noindent
{\bf Remark 12.} In this regard, it may be interesting to note that according to our model at the end of April (day 70) we will have
\beq x \ \simeq \ 0.032 \ , \ \ \ \frac{R}{R+U} \ \simeq \ 0.124 \ . \eeq
Again according to our model and fit, at the same date the fraction of individuals having gone through the infection -- and thus hopefully having acquired long-time immunity -- would however be still below 3\% nationwide; this is obviously too little to build any group immunity. The situation could be different in the areas more heavily struck by the epidemic, such as Bergamo and Brescia Departments. \EOR

\medskip\noindent
{\bf Remark 13.}
{ Needless to say, these plots and the figures given in Table \ref{tab:meas} should \emph{not} be seen as forecasts, both because we consider parameters to be constant in the future -- which will quite surely not be the case -- and more relevantly because SIR-type models do not take into account any structure (neither geographical, nor age-related, nor considering different general health status) of the population, and hence are definitely too rough to make any reliable prediction. On the other hand, we believe these models can give relevant \emph{qualitative} insights. The one emerging from our present work is that a strategy for epidemic contrast based \emph{only} on social distancing requires to keep very strong social distancing measures, possibly a general lockdown, for very long times. This is not due to details of the models considered here -- in which case the result would be just irrelevant -- but to the basic mechanism of infection propagation through contact (thus quadratic terms coupling infectives and susceptibles) and of isolation or healing through mechanisms involving a single individual (thus linear terms). \EOR
\bigskip

\begin{table}
  \centering
  \begin{tabular}{|l||c|c|c||r|r|c|c||}
  \hline
   & $\a/\a_2$ & $\b/\b_0$ & $\eta/\eta_0$ & $t_s$ & $\tau$ & $R_E$ & $x_E$ \\
  \hline
  1. & 1   & 1 & 1 & 289 & 234 & $> 5*10^5$ & 0.031 \\
  2. & 1/2 & 1 & 1 & 107 &  52 & $2.4*10^5$ & 0.019 \\
  3. & 1   & 2 & 2 &  81 &  26 & $2.4*10^5$ & 0.019 \\
  4. & 1/2 & 2 & 2 &  68 &  13 & $2.1*10^5$ & 0.010 \\
  \hline
  \end{tabular}
  \caption{Time $t_s$ of reaching the safety level $I_s$, time from adoption of new strategy to $t_s$, and total final count $R_E$ of symptomatic infections, for the different strategies listed in the text, with a factor $\s = 2$.}\label{tab:meas}
\end{table}

\subsection{Discussion}

Table \ref{tab:meas} shows, in our opinion quite clearly, that:
\begin{enumerate}

\item Continuing with present measures is simply untenable, as it would leave the country in this situation for  more than one further semester, and a large number of new symptomatic infections -- hence also of casualties -- with the ensuing continuing stress on the Hospitals system.

\item Further restrictions in the direction of social distancing would have to be kept for nearly two further months; they would be effective in reducing the number of symptomatic infectives in the future. On the other hand, as individual mobility is already severely restricted, this would basically mean closing a number of economic activities which have been considered o be priorities so far (including in the most dramatic phase), which appears quite hard on social and economic grounds.

\item A campaign of early detection could be equally effective in reducing the number of infected and hence of casualties, but would require to maintain restrictions already in place for a much shorter time, less than one month. This should be implemented through contact tracing, which does not necessarily has to go through the use of technology endangering individual freedom, as shown by the strategy used in Veneto.

\item Combining further social restrictions and early detection would reduce the number of infections to a slightly smaller figure and would need to be implemented for about two weeks. This would however meet the same problems related mentioned in item (2) above, albeit for a shorter time.

\item In all cases except if no action is undertaken, the faction of symptomatic infectives should soon fall to be between 1 and 2\%. This means we foresee a reduced stress on the Hospital system, but a continuing need to investigate, track and isolate asymptomatic infections, to avoid that at the end of restrictions they can spark a new epidemic wave.

\end{enumerate}

These considerations are of course not final; but we trust they offer a picture of what the consequences of different strategies are for what concerns strictly the epidemic dynamics. Decision makers will have of course to consider other aspects: social, economical, political, and also sanitary concerning other kind of pathologies.

\medskip\noindent
{\bf Remark 14.}  We would like to stress an important point. There are by now many papers discussing what the likely future of the COVID epidemic in this or that country should be; many of these are based on a purely statistical analysis. Here we base our discussion on a theoretical model, which takes into account the basic mechanism for the transmission of the infection elucidated above and the presence and key role of asymptomatic infectives. Statistical analysis independent of any detail for the mechanism is known to be a powerful tool, but this applies when we have a large database. For COVID we have only data ``on the fly'' -- except for the Chinese case \cite{CDC,WHOCDC}, but the data available for this have been questioned, and important corrections on key quantities (e.g. the total death toll) were done with respect to the initially estimated data.
In our opinion, it makes little sense to resort to a purely statistical analysis when the underlying database consists of a single complete experience. \EOR

\medskip\noindent
{\bf Remark 15.}  When it comes to translate the indications of the models into concrete action, the word should pass to real experts -- i.e. medical doctors working on infective pathologies. No model can describe how e.g. the use of IPD (depending moreover on how well the population is instructed on their use) impacts on the $\a$ parameter, or the like for other parameters.
Last but not least, our discussion and the predictions of the models are based on the assumption that \emph{the strategies are correctly implemented}. If e.g. a country decides to follow the strategy based on early detection, but the implementation of this is not correct (e.g. if there is not the possibility to carry out a sufficient number of rapid tests), following the right strategy will result in the wrong outcome. This is again something which cannot be discussed by a mathematical model; but it should be mentioned that not only it is important to avoid an excessive trust in such general models, but also when the conclusions are robust (as we think to be the case for the present discussion), the way in which the indications of the model are implemented can make a huge difference. \EOR

\section{Conclusions}
\label{sec:conclu}

We have considered -- in the framework of ``mean field'' epidemiological models of the SIR type, hence disregarding any structure in the population -- how different strategies aiming at reducing the impact of the epidemic perform both from the point of view of} reducing the epidemic peak and the total number of people going through the infection state, and from the point of view of reducing the time-span of the acute crisis state.

This has been discussed in general terms, both within the classical SIR model \cite{KMK,Murray,Heth,Edel,Britton} framework and with use of the recently formulated A-SIR model \cite{Gasir}, providing also some general results; in this setting, however, one deals with models with given parameters, constant in time.

On the other hand, in a real epidemic -- as the ongoing COVID one -- growing public awareness and governmental measures modify these parameters. We have considered a real case (Italy) from this point of view. After recalling that the A-SIR model considered in this paper do quite well fit the epidemiological data so far, we have discussed what would be the impact of different strategies for the near future, showing that also in this case the models predict a much shorter duration of the critical phase if further action concentrates on early detection of infectives rather than on social distancing.

\section*{Acknowledgements}

We thank L. Peliti (SMRI) and B. Golosio (Cagliari) for useful discussions. GG performed his work in lock-in at SMRI; he is also a member of GNFM-INdAM.


\end{document}